# The Anomalous Distributions and Soret Coefficient in a Nonequilibrium Colloidal System

Zhou Yanjun, Du Jiulin

*Department of Physics, School of Science, Tianjin University, Tianjin 300072, China*

**Abstract:** We study the density distribution and Soret coefficient in a nonequilibrium colloidal system by using the overdamped Langevin equation for Brownian motion in an inhomogeneous strong friction medium. Based on the relation between the temperature gradient, the interaction potential and the $q$-parameter in nonextensive statistics, we show that the colloidal particle density can be a function of the temperature and anomalously follows the noted $\alpha$-distribution, or equivalently it can also be a function of the potential energy following Tsallis distribution. With the $q$-parameter we can establish a new formula of Soret coefficient and thus bridge the gap between the ideally theoretical Soret coefficient and available experiments.

## 1. Introduction

Soret effect can exist generally in a nonequilibrium system due to interactions between the various thermodynamic forces and fluxes, where the temperature gradient can drive particle flux to lead to thermal diffusion. In the past, many systems which have been almost treated in Boltzmann–Gibbs (BG) statistics are generally extensive. When complex systems with long-rang interparticle force or mean field theories are considered, where nonextensivity holds, the BG statistics may need to be generalized. In addition to the long-range interacting systems, all nonequilibrium dissipative systems which have nonvanishing thermodynamic currents (diffusive, heat, viscous or electric), and magnetic systems appear to be nonextensive [1].

The generalization of BG statistics to nonextensive statistics (NS) has been made by introducing Tsallis entropy that depends on a nonextensive parameter $q\neq 1$ [2]. The distribution function based on Tsallis entropy is a power-law $q$-distribution. According to the present studies, NS has been proved to be very useful for the variety of nonequilibrium and complex systems [2]. In the meantime, one can find that some experimental studies of the $q$-distributions were already carried out, such as the first observation of kappa-distributions in astrophysical and space plasmas [3], the $q$-distributions in solar interior using the helioseismological measurements [4], the momentum distribution in a dissipative optical lattice [5], the anomalous diffusion in a driven-dissipative dusty plasma [6,7]. As for the colloid system, Sudo *et al* [8] found that when dynamic light scattering experiments are performed on the colloidal particles in a liquid medium by self-mixing laser Doppler velocimetry with a thin-slice solid-state laser, the power spectrum of the modulated wave induced by the motion of the colloidal particles can be described by the $q$-Gaussian distribution function. The *q*-Gaussian spectrum or, more specifically, the *q*-parameter originates from the combination of the nonlinear drag force and the local temperature fluctuation. Thus when the drag force is nonlinear and temperature of the medium fluctuates, the power spectrum of the velocity autocorrelation function of the colloidal particles would be described neither by the Lorentzian (Eq.(5) in [8]) nor by the Gaussian (Eq.(6) in [8]), but by the *q*-Gaussian (Eq.(11) in [8]).

Complex systems are often nonextensive and the statistical properties follow the power-law distributions [2]. Besides, in certain degree the physical meaning of the nonextensive parameter $q \neq 1$ has also been discussed and determined. It was proved from first principle statistical mechanics that the canonical distribution function of a system is the *q*-distribution if and only if $d(k_B T)/dE = q-1$, where $E$ is the energy of the environment (or the "heat bath") [9]. Especially, for the nonequilibrium systems with long-range interactions, one believes that the environment around the system is space inhomogeneous, rather than the uniform "heat bath" as usual, and the potential energy and the temperature are considered to be space-dependent functions. Based on the Boltzmann equation, the relation was known between the uniform temperature *T*, the interacting potential energy *U* and the *q*-parameter in NS [10,11], given by

$$k_B \nabla T = (q-1) \nabla U, \qquad (1)$$

where $k_B$ is the Boltzmann constant. For one-dimension case, Eq.(1) can be written as $q-1 = k_B (dT/dx)/(dU/dx)$. Recently, a more general expansion of Eq.(1) for rotating astrophysical systems has been developed [12], including self-gravitating systems and space plasmas.

Eq.(1) shows a relationship between the parameter $q \neq 1$ and the potential gradient (the field force) and the temperature gradient (the thermodynamic force) for the nonequilibrium many-body system with long-range interactions such as gravitational force and Coulombian force etc. This relation tell us that the parameter $q=1$ if and only if the temperature gradient is $\nabla T=0$, which gives a clear physical meaning of the *q*-parameter. The power-law *q*-distribution describes the nonequilibrium nature and nonisothermal configurations of the interacting system with potential energy *U*. For example, for the self-gravitating system with the gravitation potential $\varphi_g$, one has $\nabla U = -m \nabla \varphi_g$ [10], and in one-dimension case, $d\varphi_g/dx = GM(x)/x^2$, the nonextensive parameter can be calculated [4] by the expression,

$$1 - q = -\frac{k_B}{\mu m_H} \frac{dT}{dx} \bigg/ \frac{GM(x)}{x^2}, \qquad (2)$$

where *G* is the gravitational constant, *M(x)* is the mass interior to a sphere of radius *x*, $m_H$ is the mass of the hydrogen atom, and $\mu$ is the mean molecular weight. For the plasma with Coulombian interaction potential $\varphi_C$, one has $\nabla U = e \nabla \varphi_C$ and in one-dimension case the nonextensive parameter can be calculated [11] by

$$1 - q = \frac{k_B}{e} \frac{dT}{dx} \bigg/ \frac{d\varphi_C}{dx}, \qquad (3)$$

where *e* is the electron charge. If the plasma has magnetic field, one has $\nabla U = e(\nabla \varphi_C - c^{-1} \boldsymbol{u} \times \boldsymbol{B})$, and the nonextensive parameter [13] is expressed as

$$(1-q)\left(\nabla \varphi_C - \frac{1}{c} \boldsymbol{u} \times \boldsymbol{B}\right) = \frac{k_B}{e} \nabla T, \qquad (4)$$

where *c* is the light speed, $\boldsymbol{B}$ is the magnetic induction and $\boldsymbol{u}$ is the overall bulk velocity of the plasma. What's more, Eq.(1) has been applied to establish new characteristics of nonequilibrium complex systems, such as the plasmas with power-law distributions [14], the convection stability criterions in fluids [15,16], the phase transition of self-gravitational systems [17], the gravitational heat conduction in solar interior [18], the introduction of new physical concepts [19,20] and so on, becoming a very helpful property to study the nonequilibrium complex systems with power-law distributions.

Colloidal particles in a solution undergo Brownian motion due to the frequent collisions with the light surrounding molecules. Based on the well-controlled experiments, numerical simulations and theoretical studies [21-23], it has been known that Brownian colloidal particles obey an overdamped Langevin equation with multiplicative noise, where the total potential includes that due to the external force filed and that due to the inter-particle interactions. In a small-size magnetic colloidal particle system dispersed in a not very dilute solution, the particle core is magnetic oxide, and is coated with surface ligands in a polar solution or with a surfactant in a nonpolar solution. The stability of the colloid system is ensured by van der Waals magnetic-dipolar interactions between particles and inter-particle repulsion: electrostatic in a polar carrier and steric in a nonpolar one [24]. In order to account for the inter-particle interactions in a not very dilute solution within a one-particle framework, some kind of mean-field approach was used based on the Debye-Hückel -Onsager theory of ion transport in semi-dilute solutions of electrolytes [25]. The scope where one particle is affected by presence of nearby particles is on a scale given by the Debye length, $\lambda_D = (\varepsilon_s k_B T / 2 n_0 e^2)^{1/2}$, where $\varepsilon_s$ is the solvent permittivity and $n_0$ is the density at $x=x_0$. When at the room temperature $T$=293K, one takes $\varepsilon_s$ =78.5F/m in the water and the density $n_0$=10$^{-4}$ mol L$^{-1}$, then Debye length $\lambda_D$ is about 30 nm. At the same time, diameter $d$ of the magnetic colloidal particle can be evaluated by comparing the thermal energy with the dipole-dipole pair energy as $d \leq (72 k_B T / \pi \mu_0 M^2)^{1/3}$ [26], where $\mu_0$ is the permeability of free space and $M$ is the magnetization intensity. Usually, the particle diameter $d$ is about 10 nm. These are useful for taking into account the colloid-solvent and colloid-colloid interactions through the potential energy of one colloidal particle.

One common phenomenon for a nonequilibrium colloid system is that the concentration current can be induced by the temperature gradient and so if the initially homogeneous material is submitted to the temperature gradient, the concentration current can be generated, which is so-called Soret effect, characterized by Soret coefficient $S_T$ proportional to the volume fraction $\Phi$ or the reduced Soret coefficient $S_T^*$, defined by $S_T^* = S_T / \Phi$, independent of $\Phi$ (see Eq.(7.3) in [19]). Positive Soret coefficient means the colloids to be dragged towards lower temperatures, or sometimes it is said that the particles accumulate in regions of smaller agitation. While negative Soret coefficient means opposite [25]. As the most striking feature, in an ideal case where the interactions between the particles are neglected, Soret coefficient is $S_T^i$ =1/$T$ and always keeps positive [27]. However, it was reported that various liquid suspensions showed a change in sign as a function of external control parameters such as temperature, salinity and solute concentration, surface coating, PH value and so on [28,29]. The studies on alkali halide solutions confirmed Soret's early measurements (positive Soret coefficient) at room temperature, yet found a negative salt mobility at lower temperature. NaCl and KCl showed a sharp drop from positive to negative values as the salinity approaches 100 mMoll$^{-1}$ [28, 30]. For suspensions of colloidal silica and magnetic colloidal systems [30], both signs occur for ionic and surfacted ferrofluids, depending on the sign of the surface charge, the magnetic field, and the pH value [24, 27].

In this work, based on the Langevin equation for Brownian particles moving in an inhomogeneous strong friction medium we study the density distribution and Soret effect in a nonequilibrium colloidal system, where the temperature gradient is associated with the interacting potential by the $q$-parameter in NS. In section 2, we give a framework to derive the density distribution of the colloidal particles and the Soret coefficient. And in section 3, by using our new formula of Soret coefficient combining with the available experimental data on magnetic colloids we calculate the $q$-parameter of the system. Finally in section 4, we give the conclusion.

## 2. The anomalous distributions and Soret coefficient

In theory, the physical understanding of Soret effect is still continuing. It is worth mentioning that the kinetic theory of Brownian motion is a reliable theory for Soret problem. Firstly, in the presence of nonuniform temperature field, Kampen derived the Fokker-Planck equation for the particle density (or the particle current density) and got an additional "thermal potential" in the drift term [31], all of which cannot be obtained by directly plugging the position dependence into a homogeneous expression. Secondly, the expression of particle current density is either phenomenological or tends to minimize the free energy of the system. Such a thermodynamic rule is not applicable in the case of nonuniform temperature, and a new structure for the current is expected [25]. In view of this point, a reliable treatment of the Soret problem cannot be expected to have general validity unless it is borne out in kinetic theory. Based on well-controlled experiments, numerical simulations and theoretical studies, it has been known that Brownian colloidal particles obey an overdamped Langevin equation with multiplicative noise.

We can consider the colloidal particles in a solution to undergo Brownian motion, where the Brownian particles move in an inhomogeneous medium with strong friction and nonuniform temperature. We first start with the standard Langevin equation for the coordinate and momentum $(x, p)$ [32]:

$$\frac{dx}{dt} = \frac{p}{m}, \quad \frac{dp}{dt} = -\frac{dU(x)}{dx} - \gamma(x,p)\,p + g(x,p)\eta(t), \quad (5)$$

where $m$ is mass of the particle, $\gamma(x,p)$ is the friction coefficient, $g(x,p)$ is the noise strength and $U(x)$ is the interaction potential. Usually in normal condition, the friction coefficient and the noise strength are regarded as a constant approximately. But when the Brownian particle moves in an inhomogeneous complex medium, for generality, they both may be considered as a function of the variables $x$ and $p$, i.e. $\gamma \equiv \gamma(x,p)$, $g \equiv g(x,p)$, and the noise therefore becomes multiplicative. The white noise $\eta(t)$ is still Gaussian, with zero average and delta-correlated in time $t$, such that it satisfies,

$$\langle \eta(t) \rangle = 0, \quad \langle \eta(t)\eta(t') \rangle = 2\delta(t-t'). \quad (6)$$

And $g(x,p)$ and $\gamma(x,p)$ satisfy the fluctuation-dissipation relation: $B(x,p) \equiv g(x,p)^2 = m\gamma(x,p)\,k_B T(x,p)$.

In the strong friction case [$\gamma(x,p) \geq \gamma_0$ and $\gamma_0$ is a large number], the coordinate undergoes a creeping motion, and the derivative on momentum may be eliminated approximately from the Langevin equation (5), which leads to the following one-dimension overdamped Langevin equation only for the coordinate $x$ in the Stratonovich interpretation (see Eq.(2.20) in [33]),

$$\frac{dx}{dt} = -\frac{1}{m\gamma(x)}\frac{dU(x)}{dx} - \frac{1}{[m\gamma(x)]^2}g(x)\frac{dg(x)}{dx} + \frac{g(x)}{m\gamma(x)}\eta(t). \tag{7}$$

And the fluctuation-dissipation relation becomes $B(x) \equiv g(x)^2 = m\gamma(x)k_BT(x)$. Although there are other choices for the overdamped Langevin equations such as in the Ito and anti-Ito interpretations, the physically correct equation for the overdamped Brownian motion in an inhomogenous medium is generally Eq.(7) with the Stratonovich interpretation [23,34].

In practice, the colloidal solutions are not dilute, so the interparticle interactions should be taken into consideration. In general, the interaction potential comprises of two parts [25,27]: One comes from the interactions between the colloidal particles and the solvent, and it can be written as $U=Q\varphi/2$ [25], where $Q$ is the charge of colloids and $\varphi$ is the mean electric potential given through the surrounding solvent, based on some kinds of mean-field approach. In nonuniform temperature, the charge $Q$ may depend on the temperature owing to chemical equilibrium between the colloid surface and the solution; the other is from the colloid-colloid interaction which is sensitive to the particle density. Thus, the interaction potential is a function of the temperature and the density, i.e.

$$\frac{dU}{dx} = \left(\frac{\partial U}{\partial n}\right)_T \frac{dn}{dx} + \left(\frac{\partial U}{\partial T}\right)_n \frac{dT}{dx}. \tag{8}$$

Many authors calculated the first part of the interaction potential according to Debye-Hückel theory [25,27,35] and the second one by adopting Lennard-Jones potential [36-38] to finally give the Soret coefficient. Unlike these works, we have not relied on the concrete forms of the interaction potential but instead introduce the nonextensive parameter $q$ to connect the interaction potential and the temperature gradient. Thus, a new expression of Soret coefficient can be established.

Eq.(7) is the Langevin equation with a position-dependent multiplicative noise term. It is straightforward to utilize the Langevin equation to write the Fokker-Planck equation based on the Kramers-Moyal expansions. Following the Kramers-Moyal coefficients defined in the basic theory of Brownian motion for the Gauss-Markov process (see Eq. (1.10.14) and Eq. (1.10.17) in [39]), the drift coefficient $D^{(1)}(x)$ and diffusion coefficient $D^{(2)}(x)$ in the Fokker-Planck equation can be directly written from the Langevin equation (7) as

$$D^{(1)}(x) = -[m\gamma(x)]^{-1}\frac{dU}{dx} + \frac{B(x)}{m\gamma(x)}\frac{d}{dx}[m\gamma(x)]^{-1}, \text{ and}$$

$$D^{(2)}(x) = B(x)[m\gamma(x)]^{-2}. \tag{9}$$

If $n(x,t)$ is the particle density at the position $x$ and time $t$, corresponding to Eq.(7) and Eq.(6), the Fokker-Planck equation is written as

$$\frac{\partial n(x,t)}{\partial t} = -\frac{\partial}{\partial x}\left[D^{(1)}(x)n(x,t)\right] + \frac{\partial^2}{\partial x^2}\left[D^{(2)}(x)n(x,t)\right]. \tag{10}$$

Using Eq.(9), it can be expressed equivalently by

$$\frac{\partial n(x,t)}{\partial t} = \frac{\partial}{\partial x}\left(\frac{n(x,t)}{m\gamma(x)}\frac{dU}{dx}\right) + \frac{\partial}{\partial x}\left[[m\gamma(x)]^{-1}\frac{\partial}{\partial x}\left(\frac{B(x)n(x,t)}{m\gamma(x)}\right)\right]. \tag{11}$$

It is clear that Eq.(11) is exactly the Smoluchowski equation [39] if $\gamma$ is a constant, and thus it is a generalized Smoluchowski equation for an inhomogeneous overdamped medium. The stationary solution $n_s(x)$ of Eq. (11) satisfies the equation,

$$n_s(x)[m\gamma(x)]^{-1}\frac{dU}{dx} + [m\gamma(x)]^{-1}\frac{d}{dx}\left[\frac{B(x)n_s(x)}{m\gamma(x)}\right] = const. \tag{12}$$

And then in the zero flux boundary condition (Neumann condition) it can be solved exactly as

$$n_s(x) = \frac{C}{k_B T(x)}\exp\left(-\int\frac{dU}{k_B T(x)}\right), \tag{13}$$

where $C$ is the integral constant. In the thermal equilibrium distribution, $\ln n_s(x)$ is proportional to the ratio $U/k_B T$. At the same time, it can be seen from other papers that $\ln n_s(x)$ is usually written as the ratio, $S_T^*\Delta T$, where $\Delta T$ is the spatial temperature modulation (see Eq.(11) in [28]). In fact, both these forms are equivalent for small size nanoparticles. In the Hückel limit, the reduced Soret coefficient $S_T^*$ can be written as a temperature derivative of the charging energy [see Eq.(55) in [28], $S_T^* = -(k_B T)^{-1}dU/dT$, and thus the stationary density $\ln n_s(x)$ becomes the ratio $U/k_B T$, which coincides with the thermal equilibrium distribution. However, due to the nonuniform temperature in the nonequilibrium magnetic colloids, $\ln n_s(x)$ is not proportional to the ratio $U/k_B T$ but behaves anomalously as a function [27] such as $\int dU(x)/k_B T(x)$.

Now we discuss the forms of the integrand in Eq.(13). It is well known that if the colloidal particles moved in a homogeneous medium with uniform temperature, the density would be homogeneous. But if the colloidal particles move in an inhomogeneous medium with nonuniform temperature, the density is inhomogeneous. In the case of one dimension, taking Eq.(1) into Eq.(13), we obtain the density,

$$n_s(x) = n_0\left[\frac{T(x)}{T_0}\right]^{\frac{q}{1-q}}, \tag{14}$$

where $T_0$ is the temperature at $x=x_0$. We find that the density $n_s(x)$ is a power-law function of the nonuniform temperature and follows the noted $\alpha$-distribution with the $\alpha$-parameter $\alpha \equiv q/(q-1)$, whose dynamical origin was discussed in [32]. At the same time, we can also integrate both sides of Eq.(1) and obtain the relationship between the temperature and the interaction energy,

$$\frac{k_B}{q-1}[T(x)-T_0] = U(x) - U_0, \tag{15}$$

where $U_0$ is the potential energy at $x=x_0$. Bring Eq. (15) into Eq. (14) we find

$$n_s(x) = n_0\left[1 - (1-q)\frac{U(x)-U_0}{k_B T_0}\right]^{\frac{q}{1-q}}. \tag{16}$$

This solution possesses the form of Tsallis $q$-distribution. Hereto, a power-law form of the density distribution in the colloid system is introduced with the nonextensive parameter $q \neq 1$, which stands for a degree of deviation from thermal equilibrium distribution.

Next, we use Eq.(1) and Eq.(11) to derive the theoretical reduced Soret coefficient. Eq.(11) can be rewritten as,

$$\frac{\partial n(x,t)}{\partial t} = \frac{\partial}{\partial x}\left[\frac{k_B}{m\gamma(x)}\left(T(x)\frac{\partial n(x,t)}{\partial x} + \frac{q}{q-1}n(x,t)\frac{dT(x)}{dx}\right)\right]. \tag{17}$$

Then we make the substitution $D(x) = k_B T(x)/m\gamma(x)$, where $D(x)$ is the position-dependent diffusion coefficient. The right side of Eq.(17) is the diffusive flux comprised of the concentration gradient and the temperature gradient. Meanwhile, the diffusive flux of a dilute solution of particles of density $n(x)$ in absence of the pressure gradient is usually written (see Eq. (3.1) in [24], Eq. (1) in [27] and Eq. (58.11) in [40]) by

$$J(x,t) = -D(x)\left(\frac{\partial n(x,t)}{\partial x} + S_T^* n(x,t)\frac{dT(x)}{dx}\right). \tag{18}$$

By comparing Eq.(17) with Eq.(18) one can obtain the reduced Soret coefficient $S_T^*$,

$$T(x)S_T^* = \frac{q}{q-1}. \tag{19}$$

In Eq.(19), we have established a relation between the reduced Soret coefficient and the $q$-parameter. If $q>1$ or $q<0$, $S_T^*$ is positive, which means the colloids are dragged towards the lower temperatures, or sometimes the particles accumulate in regions of smaller agitation [25]. On the other hand, if $0<q<1$, $S_T^*$ is negative, which is opposite tendency. In the absence of the interactions, the ideal Soret coefficient follows $S_T^i = 1/T$ and comparison with available experiments shows that few systems satisfy this relation [27]. Whereas, Eq.(19) in our result bridges the gap between the theory and experiments. Lenglet et al [24] studied different types of magnetic colloids with various dilution rates and showed that the sign of the Soret effect depended little on the particle-core nature but much more on the particle surroundings (coating and nearby solvent). In our approach, the introduction of the $q$-parameter in the magnetic colloids with nonuniform temperature brings exact information about the intrinsic link between Soret effect and the interactions of the colloids with the surroundings.

## 3. Comparisons with the experiments

Using the new formula Eq.(19) we can obtain the Soret coefficient based on the $q$-parameter in Eq.(1), or we can calculate the $q$-parameter based on the Soret coefficients measured in experiments. In the following, as examples, we combine the available experimental data with Eq.(19) to calculate the $q$-parameter. The structure of the studied magnetic colloids is the same as that shown in [24]: For example, the particle core is made up of maghemite ($\gamma$-$Fe_2O_3$) or cobalt ferrite ($CoFe_2O_4$), and is denoted "207" or "184" respectively. In surfacted samples, the solvent is cyclohexane ($C_6H_{12}$) or toluene ($C_7H_8$), denoted by "CX" or "Tol" respectively; the surfactant is denoted "BNE" or "OA" for Beycostatne or oleic acid, respectively. In ionic samples, nanoparticles are dispersed in water, and colloids are stabilized by citrate ("Cit") or $H^+$ ligands, with $Na^+$ ($pH\approx7$) or $NO_3^{-1}$ ($pH\approx2$) counterions, respectively. The thermal diffusion in magnetic colloids is studied by the Forced Rayleigh scattering technique and the reduced Soret coefficient is obtained.

In Table 1, the experimental values of $S_T^*$ was listed and then using these values the $q$-parameter was calculated on the basis of Eq.(19). All the calculated $q$-parameter values are close to 1, and for CX207BNE and Tol207BNE with the same particle core and surfactant, the values of $q$ differ slightly, but for CX207BNE and CX207OA with the same particle core and solvent, there are large differences in the values of $q$. As mentioned above, the $q$-parameter brings information about the intrinsic link between the Soret effect and the interactions of the colloids with the surroundings in nonuniform temperature. The differences in the values of $q$ may be explained as differences including the interactions between the coating and the surroundings in a colloid system.

Table 1 Structure of the studied magnetic colloids and the reduced Soret coefficients [24] (Symbol "≈" before a number indicates a value estimated from nearby data, $q$ is calculated at room temperature $T$=293K).

| Sample | Structure | | | $S_T^*$ ($10^{-3}K^{-1}$) | $q$ |
|---|---|---|---|---|---|
| Surfacted | Surfactant | Core | Solvent | | |
| CX207BNE | Beycostatne | $\gamma$-$Fe_2O_3$ | $C_6H_{12}$ | 166±31 | 1.021 |
| Tol207BNE | Beycostatne | $\gamma$-$Fe_2O_3$ | $C_7H_8$ | 145±41 | 1.024 |
| CX207OA | Oleic acid | $\gamma$-$Fe_2O_3$ | $C_6H_{12}$ | 29±5 | 1.133 |
| Ionic | Stabilization | Core | Counter-ion | | |
| V207NO$_3$ | $H^+$ | $\gamma$-$Fe_2O_3$ | $NO_3^{-1}$ | −76±14 | 0.957 |
| V207Cit | Citrate | $\gamma$-$Fe_2O_3$ | $Na^{+1}$ | −185±92 | 0.982 |
| S184 | Citrate | $CoFe_2O_4$ | $Na^{+1}$ | ≈−462 | 0.993 |

For V207NO$_3$ and V207Cit with the same particle core but different coating and solvent, the values of $q$ are significantly different, but for V207Cit and S184 with different particle core but the same solvent and coating, the values of $q$ are slightly different. This may be explained as that the properties of Soret coefficient depend little on the particle-core nature [24], but the interactions between the coating and the surroundings plays a dominate role.

## 4. Conclusion

In conclusion, we have studied the density distribution of colloidal particles in a nonequilibrium colloidal system with the nonuniform temperature. We based on the overdamped Langevin equation for Brownian particles moving in an inhomogeneous strong friction medium, where the friction coefficient is position-dependent, and we employed the corresponding Fokker-Planck equation to derive the stationary density distribution of the colloidal particles. The temperature gradient is associated with the potential energy of the colloidal particles by the $q$-parameter in nonextensive statistics. We find that the colloidal density distribution can be a function of the temperature and anomalously follows the noted $\alpha$-distribution, or equivalently it can also be a function of the potential energy and follows Tsallis distribution.

Further we have studied the Soret coefficient in the nonequilibrium colloidal system. Using the $q$-parameter in NS we have established a new formula of the reduced Soret coefficient $S_T^*$. The new expression of $S_T^*$ is reasonable to bridge the gap between the theories and the available experiments.

By using the new expression of the Soret coefficient $S_T^*$ and the available experimental data on magnetic colloids as examples, we have calculated the nonextensive $q$-parameter. It is shown that the values of the $q$-parameter are slightly different from 1; the value of $S_T^*$, and even its sign, can change from one magnetic colloid to another while keeping the same core, which may prove that the Soret effect depends little on the nature of particle-cores and high much on the surroundings with nonuniform temperature. The values of $q$ can also vary from less than 1 to larger than 1, showing the interplay of the colloids with the nonuniform temperature.

**Acknowledgments**

This work is supported by the National Natural Science Foundation of China under grant No 11175128 and by the Higher School Specialized Research Fund for Doctoral Program under grant No 20110032110058.

**References**
[1] E.G.D. Cohen, Physica A **305** (2002) 19.
[2] C. Tsallis, Introduction to Nonextensive Statistical Mechanics: Approaching a Complex World, Springer, New York, 2009.
[3] V. M. Vasyliunas, J. Geophys. Res. **73** (1968) 2839.
[4] J. L. Du, Europhys. Lett. **75** (2006) 861.
[5] P. Douglas, S. Bergamini, F. Renzoni, Phys. Rev. Lett. **96** (2006) 110601.
[6] B. Liu, J. Goree, Phys. Rev. Lett. **100** (2008) 055003.
[7] B. Liu, J. Goree, Y. Feng, Phys. Rev. E **78** (2008) 046403.
[8] S. Sudo, T. Ohtomo, M. Iwamatsu, T. Osada, K. Otsuka, Applied Optics **51** (2012) 370.
[9] M. P. Almeida, Physica A **300** (2001) 424.
[10] J. L. Du, Europhys. Lett. **67** (2004) 893.
[11] J. L. Du, Phys. Lett. A **329** (2004) 262.
[12] H. N. Yu, J. L. Du, arXiv:1508.02290.
[13] H. N. Yu, J. L. Du, Ann. Phys. **350** (2014) 302.
[14] J. Gong, J. Du, Phys. Plasmas **19** (2012)063703; J. Gong, Z. Liu, J. Du, Phys. Plasmas **19** (2012) 083706; J. L. Du, Phys. Plasmas **20** (2013) 092901 and the references there in.
[15] Y. H. Zheng, Europhys. Lett. **101** (2013) 29002.
[16] Y. H. Zheng, W. Luo, Q. Li, J. Li, Europhys. Lett. **103** (2013) 10007.
[17] Y. H. Zheng, Europhys. Lett. **102** (2013) 10009.
[18] Y. H. Zheng, J. L. Du, Europhys. Lett. **105** (2014) 54002.
[19] Y. H. Zheng, J. L. Du, Europhys. Lett. **107** (2014) 60001.
[20] Y. H. Zheng, J. L. Du, Physica A **420** (2015) 41.
[21] G. Volpe, L. Helden, T. Brettschneider, J. Wehr and C. Bechinger, Phys. Rev. Lett. **104** (2010) 170602.
[22] R. Mannella and P.V. E. McClintock, Phys. Rev. Lett. **107** (2011) 078901.
[23] J.M.Sancho, Phys. Rev. E **84** (2011) 062102.
[24] J. Lenglet, A.Bourdon, J.C.Bacri, G.Demouchy, Phys. Rev. E **65** (2002) 031408.
[25] E. Bringuier, A. Bourdon, Phys. Rev. E **67** (2003) 011404.
[26] C. Scherer, Neto A. M. Figueiredo, Braz. J. Phys. **35** (2005) 718.
[27] S. Fayolle, T.Bickel, S.L.Boiteux, A.Würger, Phys. Rev. Lett. **95** (2005) 208301.
[28] A. Würger, Rep. Prog. Phys. **73** (2010) 126601.

[29] K. A. Eslahian, A. Majee, M. Maskos, A. Würger, Soft Matter **10** (2014) 1931.
[30] F. S. Gaeta, G. Perna, G. Scala, F. Bellucci, J. Phys. Chem. **86** (1982) 2967.
[31] N. G. Van Kampen, J. Phys. Chem. Solids **49** (1988) 673.
[32] J. L. Du, J. Stat. Mech. (2012) P02006.
[33] J.M.Sancho, M.S.Miguel and D. Durr, J. Stat. Phys. **28** (1982) 291.
[34] X. Durang, C.Kwon and H.Park, Phys. Rev. E **91** (2015) 062118.
[35] R. Piazza, A. Guarino, Phys. Rev. Lett. **88** (2002) 208302.
[36] P. A. Artola, B. Rousseau, Phys. Rev. Lett. **98** (2007) 125901.
[37] G. Galliéro, F. Montel, Phys. Rev. E. **78** (2008) 041203.
[38] F. Römer, F.Bresme, J.Muscatello, D.Bedeaux, J.M.Rubi, Phys.Rev.Lett. **108** (2012)105901.
[39] W. T. Coffey, Y. P. Kalmykov, J. T. Waldron, The Langevin Equation: With Applications to Stochastic Problems in Physics Chemistry and Electrical Engineering, World Scientific, Singapore, 2004.
[40] L. D. Landau, E. M. Lifshitz, Fluid Mechanics, Course of Theoretical Physics, Vol. 6, Second Edition, Butterworth-Heinemann, Oxford, 1999.